\documentclass[10pt,%reprint,
twocolumn,
 amsmath,amssymb,
 aps,
pra,
]{revtex4-2}
\usepackage[export]{adjustbox}
\usepackage{amsfonts} 
\usepackage{amsmath,amssymb}
\usepackage{bbold}
\usepackage{bm}
\usepackage[utf8]{inputenc} 
\usepackage[T1]{fontenc}    
\usepackage{graphicx}
\usepackage{hyperref}  
\usepackage{mathtools}
\usepackage{tikz}
\usepackage{times}
\usepackage{verbatim}
\usepackage{xfrac}
\usepackage[normalem]{ulem}
\usepackage{comment}
\usepackage{xcolor}
\usepackage{soul}
\usepackage{cancel}
 
\allowdisplaybreaks

\begin{document}

\title{The Emergence of the Normal Distribution in Deterministic Chaotic Maps}

\author{Dami\'an H. Zanette $^{1,2,\dagger}$ and In\'es Samengo $^{1,2,\dagger}$*}

\affiliation{{\sl $^{1}$ \quad Centro At\'omico Bariloche and Instituto Balseiro, Comisi\'on Nacional de Energía At\'omica, Universidad Nacional de Cuyo, Av. E. Bustillo 9500, 8400 San Carlos de Bariloche, R\'{\i}o Negro, Argentina.\\
$^{2}$ \quad Consejo Nacional de Investigaciones Cient\'{\i}ficas y T\'ecnicas, Argentina.}}
 
\begin{abstract}
The Central Limit Theorem states that, in the limit of a large number of terms, an appropriately scaled sum of independent random variables yields another random variable whose probability distribution tends to a stable distribution. The condition of independence, however, only holds in real systems as an approximation. To extend the theorem to more general situations, previous studies have derived a version of the Central Limit Theorem that also holds for variables that are not independent. Here, we present numerical results that characterize how convergence is attained when the variables being summed are deterministically related to one another by the recurrent application of an ergodic mapping. In all the explored cases, the convergence to the limit distribution is slower than for random sampling. Yet, the speed at which convergence is attained varies substantially from system to system, and these variations imply differences in the way information about the deterministic nature of the dynamics is progressively lost as the number of summands increases. Some of the identified factors in shaping the convergence process are the strength of mixing induced by the mapping and the shape of the marginal distribution of each variable, most particularly, the presence of divergences or fat tails.
\end{abstract}

\maketitle

\section{Introduction}

According to the central limit theorem (CLT), the sum of independent variables with finite first and second moments is governed by a Gaussian distribution when the number of summands is asymptotically large. The mean value and the variance of the Gaussian equal the sum of the individual mean values and variances, respectively. The Gaussian distribution has maximal entropy for a given variance and is reached independently of the distributions from which the summands are sampled. The convergence to the Gaussian limit, therefore, can be viewed as a loss of information about the original data. Extension to sums of variables with diverging first and second moments have been derived \cite{CLT-Levi,fat1, fat3, fat2}, the asymptotic distributions of which are no longer Gaussian but are still members of a family of so-called \emph{stable distributions}. 

Experience shows that many systems are successfully modeled by stable distributions, for example, in the theory of errors and propagation of uncertainty. This is often justified by the fact that errors, as well as many other quantities of interest, can be conceived as the sum of a large number of variables representing disparate magnitudes that appear to be unrelated. Yet,  
{ for instance, Physics dictates that all the variables describing a system of interacting particles (as opposed to an ensemble of free particles) are correlated} to one another. Therefore, the independence condition is no more than an approximation. 

To improve this approximation, extensions of the CLT have been developed also for variables that bear different degrees of statistical dependencies, including those obtained by subsequent application of a deterministic rule that produces ergodicity and aperiodicity \cite{CLT_strongmixing,CLT1,CLT1.5,CLT2,CLT2.5,CLT2.8,CLT3}. Here we analyze several systems of this type. As discussed in the next section, a conveniently modified version of the CLT exists for appropriately scaled sums of variables deterministically related to one another. Notably, the family of stable distributions for these cases coincides with the ones obtained for independent variables. These extensions provide mathematical certainty that sums of strongly correlated variables, if produced by a chaotic dynamical system, lose all memory of their original distribution, and asymptotically approach a distribution that also happens to be the limit of sums of independent variables sampled from a certain distribution. The strong statistical dependencies governing the physical world, therefore, may be legitimately ignored when describing the probability distributions of macroscopic variables, and it is legitimate to conceive the latter as a sum of a large number of microscopic, independent variables. This property greatly simplifies the description of macroscopic systems and has probably played a crucial role in the development of the theory of probability. 

In practical situations, however, it is important to know how many terms a sum needs to include for its distribution to be well described by the asymptotic result. To shed light on this question, we study in this paper the convergence of the probability distribution of a sum of perfectly correlated variables, generated by the iteration of a chaotic, deterministic map, towards the asymptotic distribution predicted by the extensions of the CLT. The aim is to characterize how the loss of information about the deterministic nature of the map depends on the number of variables that are summed together. Since previous theoretical results do not predict the rate of convergence towards asymptotic distributions in deterministic systems, our analysis is based on numerical simulations of several paradigmatic examples, and on a comparison with the behavior of randomly sampled systems with the same distributions. 

The paper is organized as follows. In Section \ref{CLT} we present the main theoretical tools to be employed later, including the extension of the CLT to variables that are strongly correlated, information-theoretical measures that quantify the differences between probability distributions, and the behavior of the variance of a sum of variables that are correlated. The following three sections apply these tools to the analysis of a chaotic dynamical system with a uniform marginal distribution and varying Lyapunov exponent (the Bernoulli map, Section \ref{sec3}), a chaotic dynamical system with a highly nonuniform marginal distribution and several types of orbits (the logistic map, Section \ref{sec4}), and an example of a process with fat-tailed distribution (Section \ref{fat}). Our main conclusions are summarized in Section \ref{conclu}.

\section{Central Limit Theorem for deterministic maps} \label{CLT}

We consider a generic one-dimensional map, $x(t+1)=f [x(t)]$, with a well-defined invariant measure $\rho_x (x)$, determined by the identity
 \begin{equation} \label{rho}
  \rho_x (x) dx= \rho_x [f(x)] df(x) =  \left[ \rho_x\circ f \right] (x)  \  f'(x) dx ,
\end{equation} 
where $ \left[\rho_x\circ  f  \right] (x)$ is the composition of functions $\rho_x (x)$ and $f(x)$, and the prime indicates differentiation with respect to $x$. We  assume that the mean value $\bar x$ is finite over the distribution $\rho_x (x)$, and --for now-- we assume the variance $\sigma_x^2$ of $x$  is also finite:
 \begin{equation} \label{valmed}
 \begin{split}
  \bar x &= \int x \rho_x (x) \mathrm{d}x <\infty,  \\
  \sigma_x^2 &= \int \left( x-\bar x \right)^2 \rho_x(x) \mathrm{d}x < \infty,     
 \end{split}
\end{equation} 
where the integrals run over the whole domain of variation of $x$. In Section \ref{fat} we study a case where we relax the condition that $\sigma_x^2$ is finite. A central-limit theorem (CLT) for this kind of system applies \cite{CLT1,CLT1.5,CLT2,CLT2.5,CLT2.8,CLT3} when the map under study is ergodic and aperiodic. We recall that a map is ergodic if all its invariant sets are null or co-null, and aperiodic if its periodic orbits form a null set \cite{CLT3}. The combination of ergodicity and aperiodicity is typically equivalent to the dynamics being chaotic \cite{chaos}. In this case, the CLT states that the distribution of the (centered, suitably normalized) sums of $N$ successive values of $x(t)$, 
 \begin{equation} \label{sN}
  s_N (t) =\frac{1}{\sqrt{N}} \sum_{k=1}^N \left[  x(t+k-1) - \bar x \right],
\end{equation} 
becomes normal for $N\to \infty$:
 \begin{equation} \label{normal}
\rho_s (s) =\frac{1}{\sqrt{2\pi \sigma_s^2}} \exp\left( -\frac{s^2}{2 \sigma_s^2}\right) \equiv G_{\sigma_s} (s),
\end{equation} 
for some value of the variance $ \sigma_s^2$. Here, $G_{\sigma_s}$ denotes the Gaussian centered at zero, with standard deviation $\sigma_s$.

For each value of $N$, the variables $x(t)$ and $s_N(t)$ can be integrated into a single two-dimensional map:
 \begin{equation} \label{2dimmap}
\left\{ 
\begin{array}{lllll}
     x(t+1) &= & f [x(t)], & \\
     s_N(t+1)& = & s_N(t)  & - & \frac{1}{\sqrt{N}} x(t)+ \\
     & &  & + & \frac{1}{\sqrt{N}}  f^{(N)} [x(t)],
\end{array}
\right.
\end{equation} 
where $f^{(N)} (x)= \overbrace{f \circ f \circ \cdots \circ f}^{N}(x)$ is the $N$-th self-composition of $f(x)$. Thus, for $N\to \infty$, the marginal invariant measures of the variables $x$ and $s_N$ in map (\ref{2dimmap}) are, respectively, $\rho_x (x)$ and the Gaussian $\rho_s (s)=G_{\sigma_s} (s)$ of Equation (\ref{normal}).

In contrast with the sums of statistically independent random variables drawn from a given distribution, in the limit $N\to \infty$, the variance of the sums $s_N(t)$  does not necessarily coincide with that of the summands, $\sigma^2_x$. The difference arises from the correlations between successive values of $x(t)$, induced by the map $x(t+1)=f[x(t)]$, with the ensuing mutual correlations between the values of $s_N(t)$. For a finite number of summands $N$, the variance of $s_N(t)$ is given by  the Green-Kubo formula \cite{Wouters}:
 \begin{align} 
       \sigma_{s_N}^2 = & \sigma_x^2 + \label{GK1} \\
       & + 2 \sum_{k=1}^{N-1} \left( 1-\frac{k}{N} \right) \overline{[x(t)-\bar x][x(t+k)-\bar x]}, \nonumber
\end{align} 
where the overline indicates the average with respect to the distribution $\rho_x (x)$. The value of $\sigma_{s_N}^2$ becomes independent of $t$ when the process $x(t)$ has reached a stationary regime. For $N\to \infty$, the variance is
 \begin{equation} \label{GK2}
 \begin{split}
\sigma_{s}^2 &= \lim_{N\to \infty} \sigma_{s_N}^2  \\
& = \sigma_x^2 + 2 \sum_{k=1}^\infty \overline{[x(t)-\bar x][x(t+k)-\bar x]}.     
 \end{split}
\end{equation} 
Provided that the sum converges, this formula gives the variance of the asymptotic normal distribution $G_{\sigma_s} (s)$ of increasingly long sums $s_N(t)$.

In the following, we study the process of convergence towards the asymptotic distribution predicted by the above CLT for some selected deterministic maps, as the number of terms in the sums $s_N$ grows. For each $N$, we numerically iterate  Equations (\ref{2dimmap}) and estimate the distribution of the sums $s_N$, $\rho_{s_N} (s_N)$, as a suitably normalized $10^3$-column histogram built from, typically, $10^7$ values of $s_N$. To quantify the difference between $\rho_{s_N}$ and the expected asymptotic Gaussian distribution $G_{\sigma_s} (s)$, we use the Kullback-Leibler divergence (KLD). Recall that the KLD between two distributions $\rho_1 (s)$ and $\rho_2 (s)$ is defined as 
 \begin{equation} \label{KL}
  D \left(\rho_1||\rho_2\right) = \int \rho_1 (s)  \log_2 \left[ \frac{\rho_1 (s) }{\rho_2 (s) }\right] ds.
\end{equation} 
This quantity measures the inefficiency with which the data $s$ is represented by a code optimized to be maximally compact under the assumption that the distribution is $\rho_2$ when, in reality, the data are generated from $\rho_1$. The inefficiency equals the mean number of extra bits per sample \cite{cover1992}. The divergence only vanishes when the two distributions coincide, and is otherwise positive. For brevity, we hereafter denote as $D_G$ the KLD between the distribution $\rho_{s_N}$ and the asymptotic normal distribution $G_{\sigma_s}$: $D_G \equiv D\left(  \rho_{s_N} || G_{\sigma_s} \right)$.

Additionally, for each $N$, it is interesting to compare $\rho_{s_N}$ with a normal distribution with the variance $\sigma_{s_N}^2$ given by Equation (\ref{GK1}), namely, the same variance as the sums $s_N$. Since $\sigma_{s_N}^2 \to \sigma_s^2$ as $N\to \infty$, this is an alternative way of characterizing the convergence to the asymptotic Gaussian $G_{\sigma_{s}}$. For this comparison, we introduce the KLD $D_{G_N} \equiv D\left( \rho_{s_N} || G_{\sigma_{s_N}} \right)$.

Finally, in order to contrast the deterministic dynamics of the chaotic map under study with a genuinely aleatory process, we calculate the KLD for the distribution of sums of the same form as in Equation (\ref{sN}), but with the $N$ values of the variable $x$ drawn at random from the invariant measure $\rho_x (x)$. According to the standard CLT for statistically independent variables, as $N$ grows, the distribution $\rho_{s_N}^{\rm random}$ of these random-sampling sums is expected to asymptotically converge to a Gaussian with variance $\sigma_x^2$. To quantify this convergence, we compute $D_{\rm random} \equiv D\left( \rho_{s_N}^{\rm random} || G_{\sigma_x} \right)$. 

The measures $D_{\rm random}$, $D_G$ and $D_{G_N}$ reflect three different aspects of the convergence of $\rho_{s_N}$ to $G_{\sigma_s}$. The process by which $D_{\rm random}$ tends to zero describes how independent variables, when summed together, lose the memory of the distribution from which they are sampled and approach a Gaussian. The Gaussian distribution is the one with maximal entropy among those with fixed variance. By acquiring a Gaussian shape, therefore, the distribution of the sum maximizes uncertainty. In Appendix \ref{appB}, we show that for large $N$ the divergence $D_{\rm random}$ decays as $N^{-1}$ if $\rho_x$ is not symmetric around its mean value, and at least as fast as $N^{-2}$ if there is symmetry.

A steep decay of $D_{G_N}$ with $N$, at a faster rate than $D_G$, implies a rapid evolution of $\rho_{s_N}$ towards a bell-shaped distribution, whose variance may still have to evolve to its asymptotic value $\sigma_s^2$. The convergence process can therefore be conceived as a sequence of two stages, the first one consisting of shedding all the structure in $\rho(x)$ and becoming Gaussian-like, and the second, adjusting the variance. Once $\rho_{s_N}$ is approximately Gaussian, its KLD with the asymptotic distribution $G_{\sigma_s}$ can be analytically calculated in terms of their respective variances:
\begin{equation} \label{eq:dklgauss}
D_G \approx \log_2 \left(\frac{\sigma_{s_N}}{\sigma_{s}} \right) + \frac{1}{2 \ \ln 2} \ \frac{\sigma^2_s - \sigma^2_{s_N}}{\sigma^2_{s_N}}.
\end{equation}

\section{The Bernoulli map} \label{sec3}

As a first case of study, we take the generalized Bernoulli map
 \begin{equation} \label{Bernoulli}
 x(t+1) =f[x(t)] = \{ m x(t) \},
\end{equation} 
where $\{ \cdot \}$ indicates fractional part, and $m>1$ is an integer factor.  This map has been extensively studied since long ago as a paradigm of deterministic chaotic systems, due to its combination of complex behavior and analytical traceability. Its Lyapunov exponent equals $\ln m$. The invariant measure of $x(t)$ is particularly simple:
 \begin{equation} \label{rhoB}
\rho_x(x) = \left\{ 
\begin{array}{ll}
     1&  \mbox{for $x\in [0,1)$},\\
     0& \mbox{otherwise} ,
\end{array}
\right.
\end{equation} 
for all $m$, with $\bar x=1/2$ and $\sigma_x^2=1/12$. We show in Appendix \ref{appA} that the variances of the sums $s_N$ can be explicitly calculated:
 \begin{equation} \sigma_{s_N}^2 = \frac{1}{12} + \frac{1}{6(m-1)} \left( 1-\frac{m}{m-1} \frac{1-m^{-N}}{N}\right) . \label{sigmasB}
\end{equation} 
Note that for $N\gg (\ln m)^{-1}$, this variance takes the approximate form
 \begin{equation} \label{eq:sigdeene}
\sigma_{s_N}^2 \approx \frac{1}{12}  \frac{m + 1}{m - 1} \left(1 - \frac{N_0}{N} \right),
\end{equation} 
with $N_0=2m/(m^2-1)$. For $N\to \infty$, in turn,
 \begin{equation}
\sigma_{s_N}^2\to \sigma_{s}^2 =  \frac{1}{12}  \frac{m+1}{m-1}.
\end{equation} 

\begin{figure*}[ht]
\includegraphics[width=10cm]{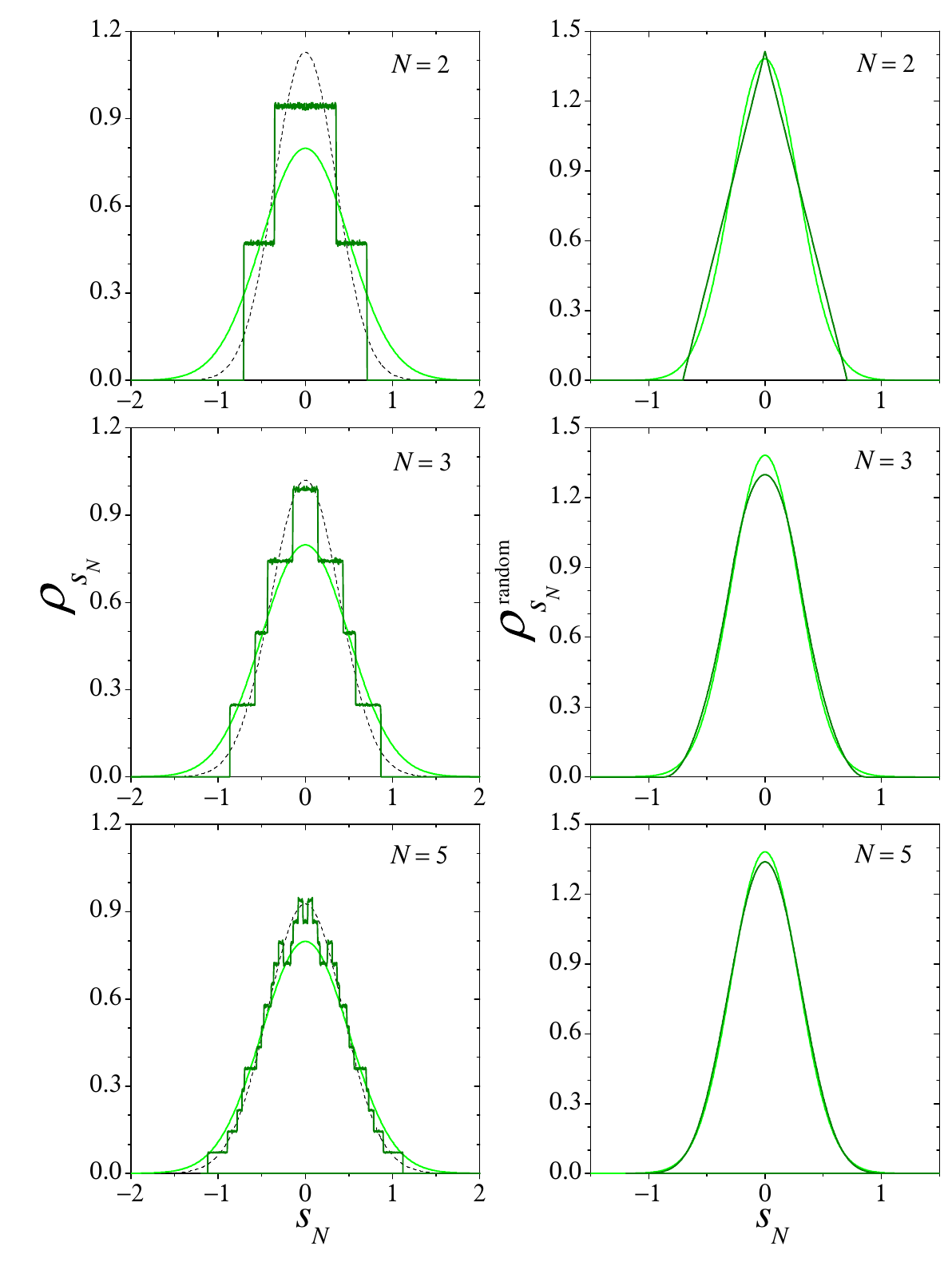}
\caption{Left, dark line: Numerical results for the distribution $\rho_{s_N}$ of the sums $s_N$ defined in Equation (\ref{sN}), in the case the Bernoulli (\ref{Bernoulli}) map with $m=2$, for three small values of $N$. The light curve is the Gaussian expected for $N\to \infty$, and the dashed curve is a Gaussian with the same variance as predicted for $\rho_{s_N}$. Right, dark curve: The distribution $\rho_{s_N}^{\rm random}$ for sums of $N$ values of $x$ randomly sampled from $\rho_x(x)$ is a normalized version of the Irwin-Hall distribution \cite{Irwin1927,Hall1927}, which can be obtained analytically by successive self-convolution of $\rho_x$. The light curve is the Gaussian expected for $N\to \infty$.  Note different scales on the left and right columns. \label{fig01}} 
\end{figure*} 

We first consider the Bernoulli map for $m=2$. Dark full lines on the left column of Figure \ref{fig01} show numerical results for the distributions of the sums $s_N$, $\rho_{s_N}$, for three small values of $N$. Light-colored curves stand for the asymptotic Gaussian $\rho_s=G_{\sigma_s}$, and dashed curves are the Gaussians $G_{\sigma_{s_N}}$ for each $N$. Their respective variances, $\sigma_s^2$ and $\sigma_{s_N}^2$, are given by Equation (\ref{sigmasB}). On the right column, dark and light-colored curves respectively show the distributions of the sums of randomly sampled values of $x$, $\rho_{s_N}^{\rm random}$, calculated analytically as $N$-th order self-convolutions of $\rho_x(x)$, and the expected asymptotic Gaussian $G_{\sigma_x}$. A comparison of the two columns illustrates the difference between the distributions of the sums generated by map iteration on one side and by random sampling on the other. It also shows that convergence to the asymptotic distribution is faster in the latter case. 

\begin{figure}[ht]
\includegraphics[width=8 cm]{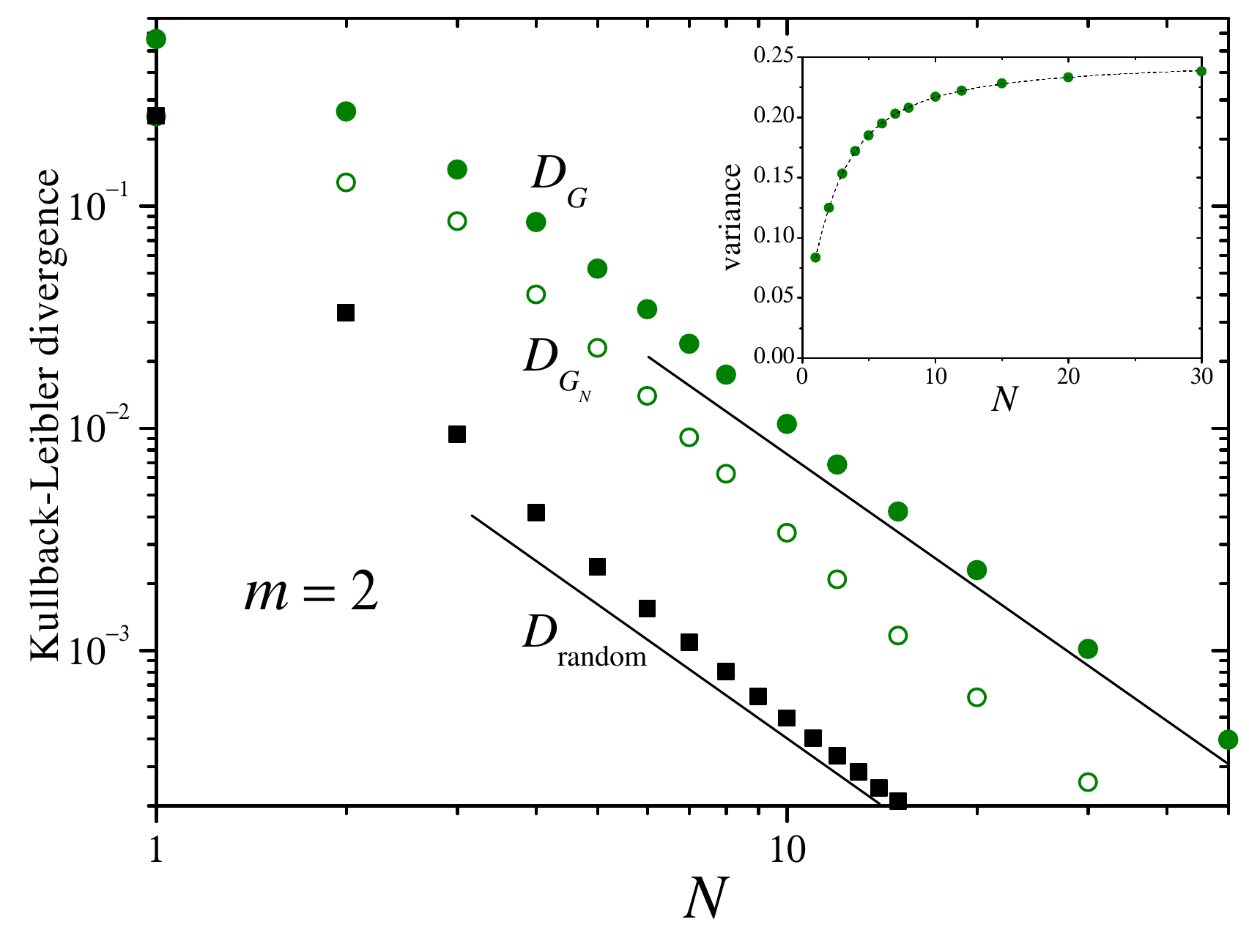}
\caption{Main panel: The Kullback-Leibler divergences $D_G$, $D_{G_N}$, and $D_{\rm random}$, defined in the text, as functions of the number of terms in the sums $s_N$ of Equation (\ref{sN}), for the Bernoulli map (\ref{Bernoulli}) with $m=2$. The straight lines in this log-log plot have a slope $-2$. The inset shows, as dots, numerical results for the variance $\sigma_{s_N}^2$ over the distribution $\rho_{s_N} (s_N)$. The dashed line joins the analytical values predicted from Equation (\ref{sigmasB}).\label{fig02}}
\end{figure} 

The main panel of Figure \ref{fig02} shows, with different symbols, the KLDs $D_G$, $D_{G_N}$, and $D_{\rm random}$, defined in the preceding section. For $N=1$, by definition, $D_{\rm random} = D_{G_N}$. For large $N$ (Appendix \ref{appB}), $D_{\rm random}$ decays as $N^{-2}$. The straight lines in the log-log plot of the figure have slope $-2$, suggesting that the decay of the divergence $D_G$ approximately follows the same asymptotic dependence on $N$. The inset in Figure \ref{fig02} shows, as dots, the numerical estimation of the variance of $s_N$ over the distribution  $\rho_{s_N}$ as a function of $N$. The dashed curve corresponds to the analytical expression of Equation (\ref{sigmasB}). 

In the range shown in the figure, for $N\gtrsim 10$,  $D_G$ is larger than $D_{\rm random}$ by a factor of around $14$. Meanwhile, in the same range, $D_{G_N}$ decays faster, approximately as $N^{-2.3}$. As discussed at the end of Section \ref{CLT}, this faster decay of $D_{G_N}$ suggests that $\rho_{s_N}$ is rapidly approaching a Gaussian distribution, with a KLD with the asymptotic distribution $\rho_S$ as given by Equation (\ref{eq:dklgauss}). Replacing Equation (\ref{eq:sigdeene}) into Equation (\ref{eq:dklgauss}) and expanding up to second order in $N_0/N$ yields

\begin{equation} \label{eq:predicciongaussiana}
D_G \approx \frac{1}{4\ln 2} \ \frac{N_0^2}{N^2}.
\end{equation}
For $m = 2$ we have $N_0 =4/3$ so that, according to the above equation, $D_G \approx 0.64 \ N^{-2}$. A power-law fitting of the data for $D_G$ for $N\le 20 \le 50$ gives $D_G \approx 0.69 \ N^{-1.9}$, which fits the prediction of Equation (\ref{eq:predicciongaussiana}) remarkably well. This agreement provides strong evidence in favor of the hypothesis that $\rho_{s_N}$ converges to $\rho_s$ in two stages, acquiring a Gaussian shape in the first, and adjusting the variance in the second. The transition from the first stage to the second, however, does not imply that $\rho_{s_N}$ is strictly speaking a Gaussian distribution.  

What are the implications of the fact that after the initial transient $D_G$ and $D_{\mathrm{random}}$ both decay with the same power law, approximately proportional to  $N^{-2}$? In this regime, $D_G \approx 14 \ D_{\mathrm{random}}$ which means that, for each $N$, $D_{\mathrm{random}}(N)$ is approximately equal to $D_G (\sqrt{14} N)$. By increasing the number of random samples drawn from the invariant measure (\ref{rhoB}), $D_{\mathrm{random}}$ diminishes by the same amount as $D_G$ diminishes when running the Bernoulli deterministic mapping a rescaled, larger number of samples, with a scaling factor of $\alpha\approx \sqrt{14} \approx 3.7$. In other words, $\alpha$ samples of the deterministic map are as informative about the asymptotic distribution as a single sample in the random drawing. The presence of correlations makes each new sample from the deterministic dynamics less informative (by a factor of $\alpha$) than from purely independent draws.

The factor $\alpha$ may also be semi-quantitatively associated with the relation between the asymptotic variance $\sigma^2_s$ and the original variance $\sigma^2_x$. In Equation (\ref{sN}), the normalization factor $1/\sqrt{N}$ compensates for the fact that the variance of a sum of $N$ independent samples is proportional to $N$. Yet, when the summands bear statistical interdependence, the intended compensation need not be attained. The higher the correlations in the deterministic map, the less informative each new datum is, the more unsuccessful the compensation, and the larger the increase of the asymptotic variance. In the present case, the variance increases threefold, from $1/12$ to $1/4$, which is similar to the factor relating $D_G$ and $D_{\mathrm{random}}$, namely, $\alpha$.  
\begin{figure}[htbp]
\includegraphics[width=8 cm]{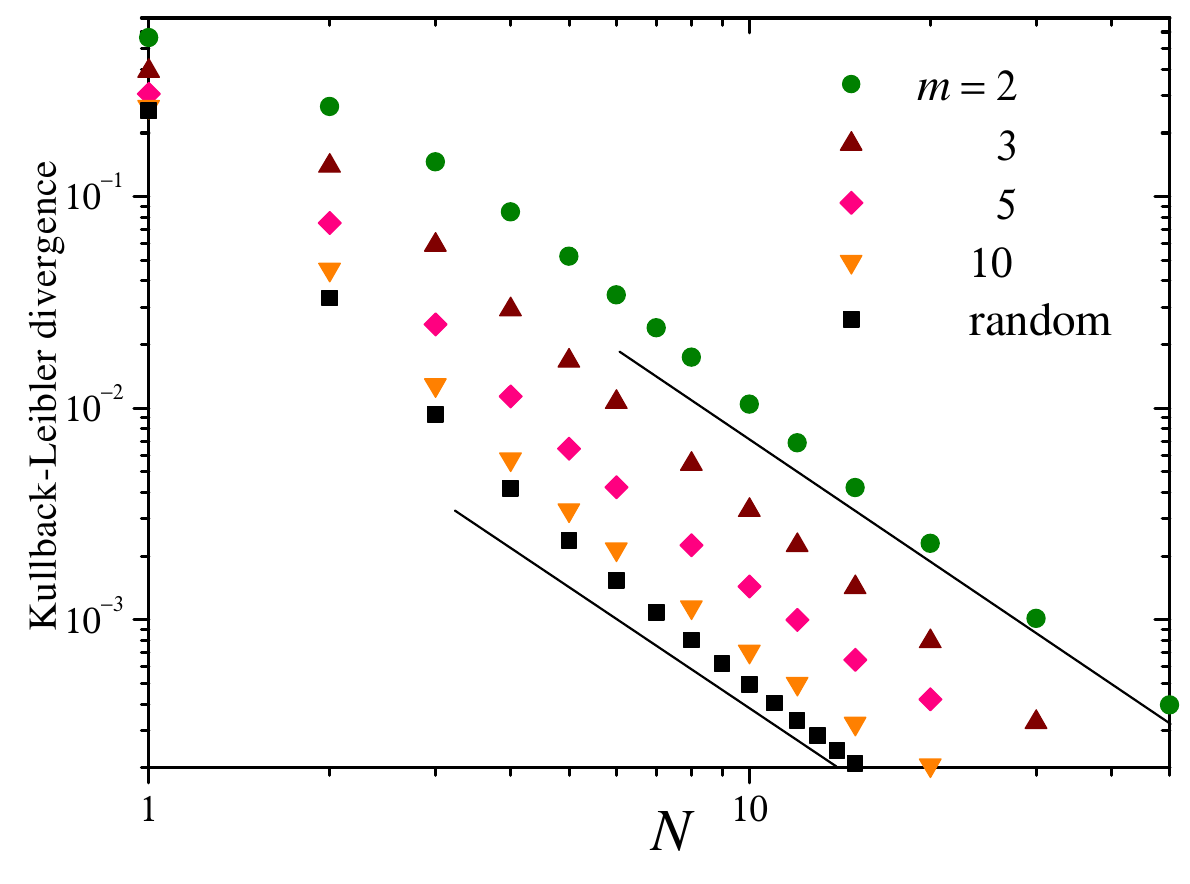}
\caption{The Kullback-Leibler divergence $D_G$ for the Bernoulli map (\ref{Bernoulli}) with various values of $m$, and $D_{\rm random}$ (which is the same for all $m$). The straight lines have slope $-2$.  \label{fig03}} 
\end{figure} 

Considering now other values of $m$ in the Bernoulli map (\ref{Bernoulli}), the numerical results presented in Figure \ref{fig03} show that the dependence of $D_G$ on $N$ is similar to that obtained for $m=2$, with the only difference that $D_G$ becomes progressively smaller as $m$ grows. As before, the convergence may be conceived as consisting of two stages, with  Equation (\ref{eq:dklgauss}) approximately holding for the second stage. According to the results of Figure \ref{fig03}, the second state is reached faster for larger values of $m$. As expected from the large-$N$ asymptotic behavior of $D_G$ predicted by Equation (\ref{eq:predicciongaussiana}) with $N_0=2m/(m^2-1)$ [cf. Equation (\ref{eq:sigdeene})], it approaches $D_{\rm random}$ for large $m$. This implies that  the effect of the statistical dependencies induced by the deterministic nature of the map decreases as $m$ grows. The KLD $D_{G_N}$ is not shown in Figure \ref{fig03}, but its behavior is similar to that of the case of $m=2$ (Figure \ref{fig02}). 

In summary, in the Bernoulli map, $D_{G_N}$ decreases faster than $D_G$ during the first stage of the convergence process, where $\rho_{s_N}$ acquires a Gaussian-like shape. Only later, the variance is adjusted towards its final value. The second stage can be modeled analytically, providing a good qualitative description of the asymptotic behavior inferred from numerical results.

\section{The logistic map: full chaos and intermittency} \label{sec4}

We now turn our attention to the logistic map \cite{May1976,Tsuchiya1997}
 \begin{equation} \label{logistic}
 x(t+1) =f[x(t)] =\lambda x(t) [1-x(t)],
\end{equation} 
with $0<\lambda \le 4$. Much like Bernoulli's, the logistic map hardly needs any presentation. We first consider the case $\lambda=4$, which we call the regime of ``full chaos''.  For this value of $\lambda$, the dynamics are chaotic and therefore comply with the hypotheses of the CLT for deterministic systems discussed in Section \ref{CLT}. Moreover, due to the existence of a nonlinear change of variables that transforms the logistic map with $\lambda=4$ into the Bernoulli map of Equation (\ref{Bernoulli}) with $m=2$, several analytical results for the latter can be extended to the former. In spite of this connection, as we show below, the statistics of the sums $s_N$ are qualitatively different between the two maps.  

For $\lambda=4$, the invariant measure of the logistic map can be written explicitly as \cite{Jakobson1981}
 
\begin{equation} \label{rhologistic}
 \rho_x (x) = \frac{1}{\pi \sqrt{x(1-x)}}
\end{equation} 

for $0\le x\le 1$, and $0$ otherwise. The mean value is $\bar x = 1/2$ and the variance is $\sigma_x^2=1/8$. As we show in Appendix \ref{appA}, the correlations between iterations of the map, $c_k=\overline{[x(t)-\bar x][x(t+k)-\bar x]}$, vanish for all $k$. From Eqs.~(\ref{GK1}) and (\ref{GK2}), this implies that the variances of the sums $s_N$ are the same for all $N$, and therefore coincide with both the variance of $x$ and with the limit for $N\to \infty$: $\sigma_{s_N}^2 = \sigma_{s}^2=\sigma_x^2$. Therefore --in contrast with the Bernoulli map studied in the preceding section-- 
it is not possible to discern between a first stage of convergence to a Gaussian profile, and a second stage of adjustment of the variance.

\begin{figure*}[ht]
\includegraphics[width=10 cm]{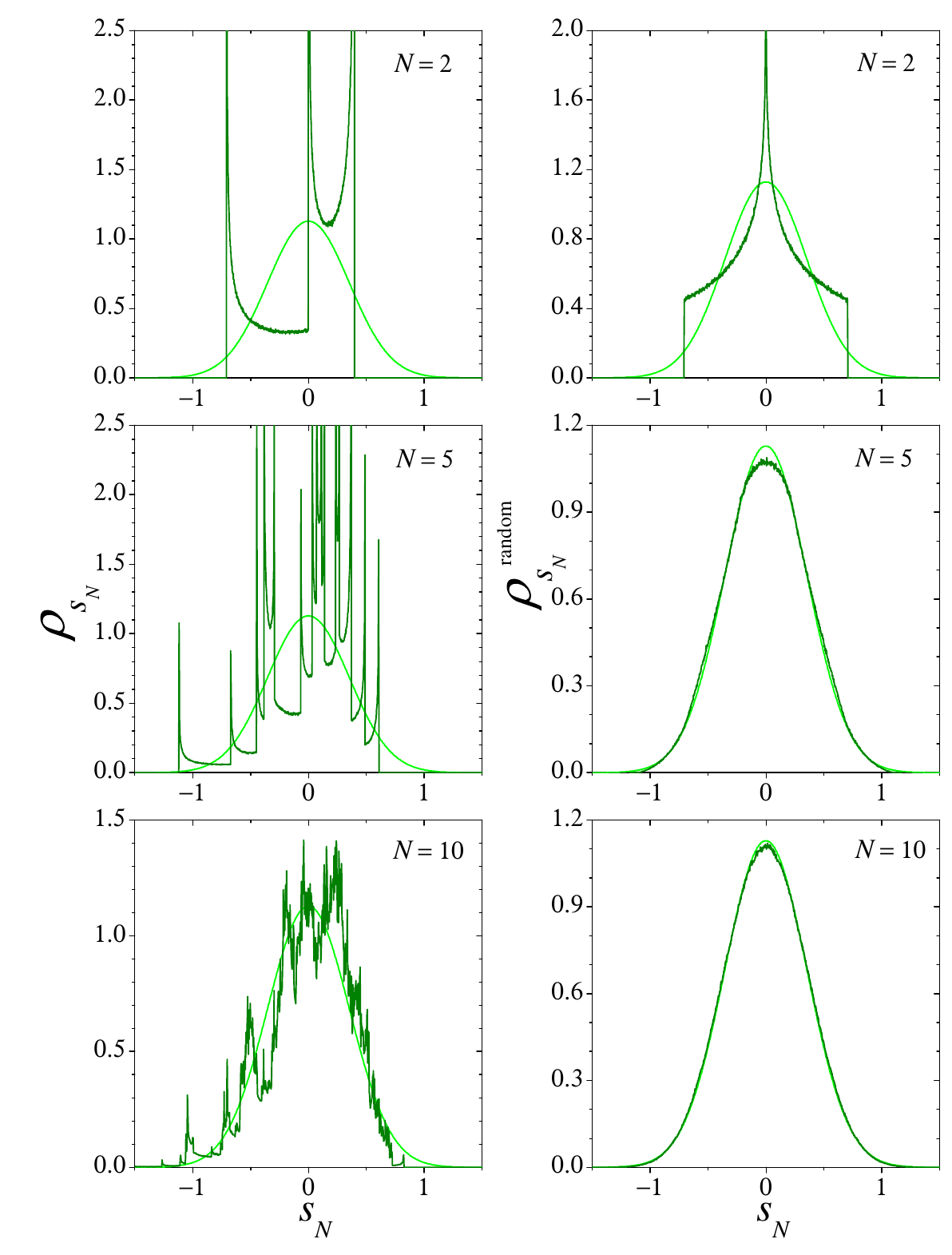}
\caption{As in Figure \ref{fig01} for the logistic map of Equation (\ref{logistic}) in the regime of full chaos, $\lambda=4$. The distributions $\rho_{s_N}^{\rm random}$, dark lines on the right column, have now been obtained numerically. Note the different scales in different panels. \label{fig04}} 
\end{figure*}

In Figure \ref{fig04}, the left column shows numerical estimations of the distributions $\rho_{s_N} (s_N)$ of the sums of $N$ consecutive iteration of the logistic map with $\lambda=4$, for three values of $N$. The light-colored curve corresponds to the expected asymptotic Gaussian. 

In addition to the sharp peaks in the profile of $\rho_{s_N}$ for small $N$, an important difference with the Bernoulli map (Figure~\ref{fig01}) is that $\rho_{s_N}$ is no longer symmetric with respect to zero. This asymmetry may come as a surprise, taking into account that both $f(x)$ and $\rho_x(x)$ are symmetric around the mean value $\bar x$. The asymmetry, however, originates from the fact that the functions $x+f(x)$, $x+f(x)+f^{(2)}(x)$, $x+f(x)+f^{(2)}(x)+f^{(3)}(x)$, $\dots$, which ultimately determine the distributions of the sums $s_N$, are {\em not} symmetric around $\bar x$.

On the right column of Figure \ref{fig04} we show, for the same values of $N$, the distributions $\rho_{s_N}^{\rm random}$ of sums of $N$ random values of $x$ sampled from $\rho_x$. In contrast with the case of the Bernoulli map, $\rho_{s_N}^{\rm random}$ is here estimated numerically. As expected, the distributions of random-sampling sums are now symmetric with respect to zero, and exhibit a fast convergence to the asymptotic Gaussian.

\begin{figure}[ht]
\includegraphics[width=8 cm]{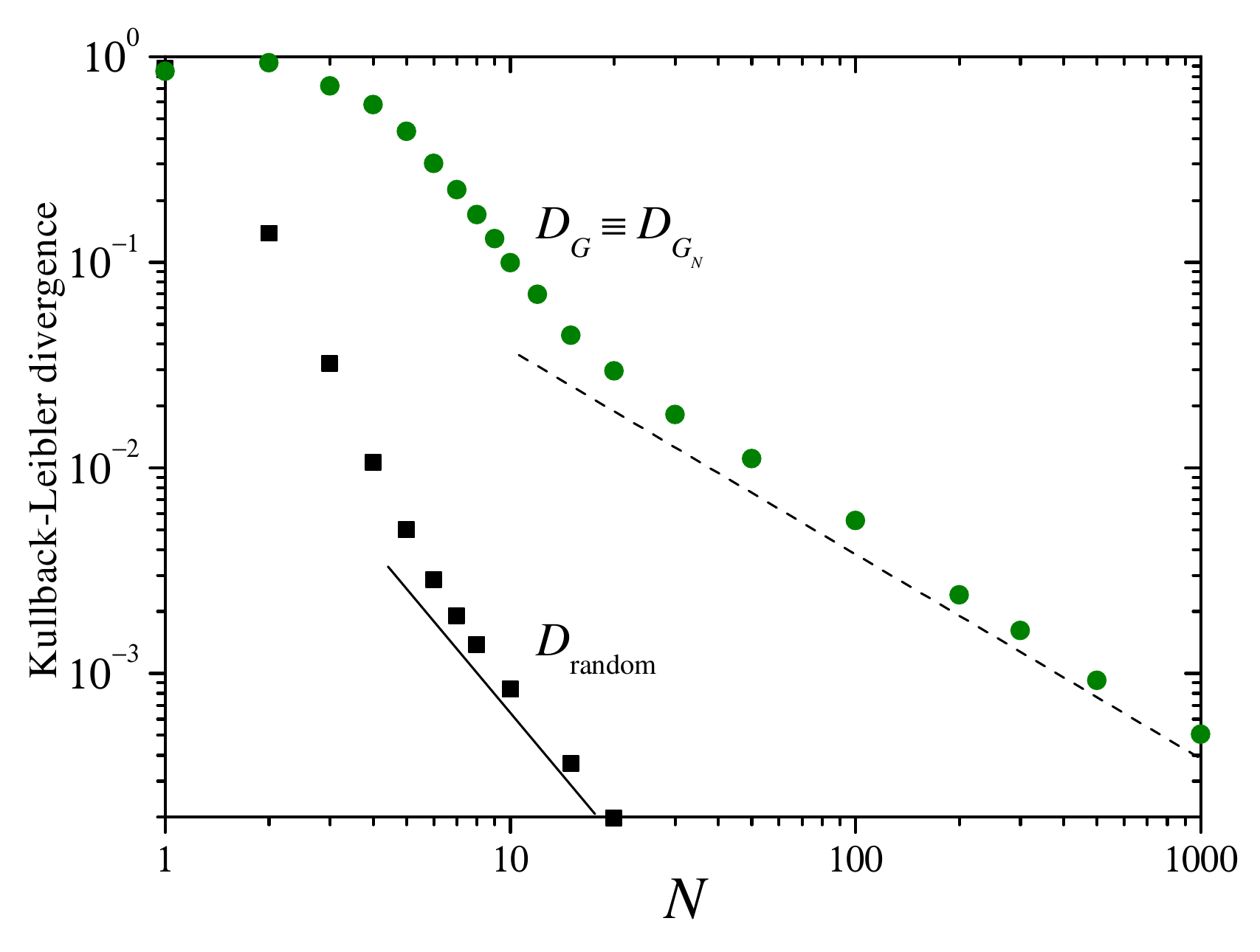}
\caption{The Kullback-Leibler divergence $D_G$ for the logistic map of Equation (\ref{logistic}) in the regime of full chaos, $\lambda=4$, and $D_{\rm random}$, as a functions of $N$. In this case, $D_{G_N}$ coincides with $D_G$. The full and dashed straight lines have slopes $-2$ and $-1$, respectively. \label{fig05}} 
\end{figure}

Figure \ref{fig05} shows $D_G$ and $D_{\rm random}$ for the fully chaotic logistic map, as functions of $N$. Since, as explained above, $\sigma_{s_N}^2$ equals $\sigma_s^2$ for all $N$, now $D_{G_N}$ coincides with $D_G$. Due to the symmetry of $\rho_x$ with respect to its mean value, the arguments given in Appendix \ref{appB} apply to this case, and $D_{\rm random}$ decays as $N^{-2}$ for large $N$. The full straight line in the log-log plot of the figure has slope $-2$, confirming this prediction in the plotted range. Yet, the behavior of $D_G$ is considerably different. It starts with a small increment between $N=1$ and $2$, where it attains a maximum, and thereafter decays rapidly up to $N\approx 20$. This decay corresponds to the interval of $N$ for which the distribution $\rho_{s_N}$ displays identifiable singularities. For $N\gtrsim 20$, the singularities start to overlap, and the distribution $\rho_{s_N}$ varies more smoothly and displays a well-defined asymmetric bell-shaped profile. In this zone, the decay of $D_G$ is slower and approximately behaves as $N^{-1}$, as illustrated by the dashed straight segment of slope $-1$.  As shown in Appendix \ref{appB}, a decay as $N^{-1}$ is expected for the KLD of the distribution of random-sampling sums when the distribution of the individual summands is not symmetric with respect to the mean value. If the disparate dependence on $N$ between $D_G$ and $D_{\rm random}$ persists as $N$ grows beyond the range considered here, their relative difference would increase indefinitely for $N\to \infty$.

\begin{figure*}[ht]
\includegraphics[width=\textwidth]{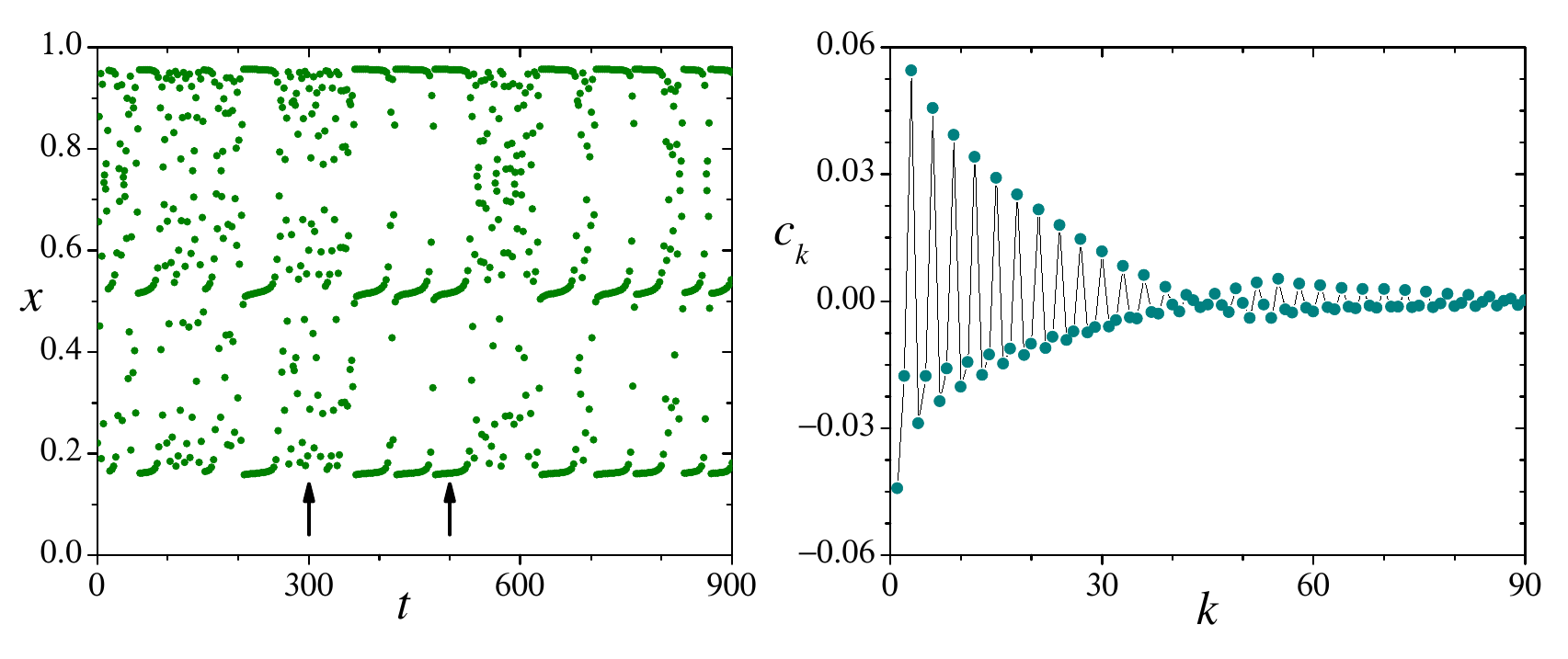}
\caption{Left: $900$ successive iterations of the logistic map, Equation (\ref{logistic}), in the intermittent regime, $\lambda=3.828$. The arrows at $t=300$ and $500$ point at ``turbulent'' and period-$3$ ``laminar'' intervals, respectively. Right: The correlation  $c_k=  \overline{[x(t)-\bar x][x(t+k)-\bar x]}$ as a function of $k$ in the same intermittent regime, calculated numerically from sequences of $10^7$ iterations of $x(t)$.  Symbols are connected by lines to facilitate visualization.\label{fig06}} 
\end{figure*}

Although still chaotic, other values of $\lambda$ in Equation (\ref{logistic}) give rise to qualitatively different dynamical features in the logistic map. For $\lambda=3.828$, which is our next case of study, the dynamics are intermittent. Just above this value of $\lambda$, at $\lambda_3=1+2\sqrt{2} \approx 3.8284$,  the logistic map enters the largest stability window within its chaotic regime, where $x(t)$ becomes asymptotically locked in a period-$3$ orbit. For $\lambda\lesssim \lambda_3$, the vicinity of the critical point manifests itself in the form of intermittent behavior for $x(t)$. Namely, the dynamics alternate intermittently between intervals of ``turbulent'' evolution, where its behavior is conspicuously chaotic, and ``laminar'' evolution, where $x(t)$ remains temporarily close to the period-$3$ orbit, but eventually departs away from it. The left panel of Figure \ref{fig06} shows $900$ successive iterations of $x(t)$ for the above value of $\lambda$, illustrating both kinds of behavior. 

For $\lambda=3.828$,  no analytical description of the logistic map exists, and we must resort to numerical techniques. As inferred from the left panel of Figure \ref{fig06}, in this case, $\rho_x (x)$ covers only a portion of the interval $[0,1]$, between $x \approx 0.157$ and $0.957$, and displays three peaks near the values of $x$ in the period-3 orbit. Our numerical estimations for the mean value and the variance are $\bar x \approx 0.593$ and $\sigma_x^2 \approx 0.0864$. In principle, the variance of the sums $s_N$ could be obtained from Equations (\ref{GK1}) and (\ref{GK2})  by numerically computing the correlations $c_k=  \overline{[x(t)-\bar x][x(t+k)-\bar x]}$. These quantities, however, exhibit sharp oscillations and slow convergence as $k$ grows, as well as persistent fluctuations for large $k$. The right panel in Figure \ref{fig06} shows $c_k$ up to $k=90$. In practice, such features make impossible the evaluation of the variances $\sigma_{s_N}^2$ and $\sigma_s^2$ using the sums in Equations (\ref{GK1}) and (\ref{GK2}). We therefore resort to their direct numerical calculation using the values of $s_N (t)$ obtained from successive map iterations. In particular, our estimation for the variance of the sums in the limit $N\to \infty$ is $\sigma_s^2 \approx 0.0403$. 

\begin{figure}[ht]
\includegraphics[width=8 cm]{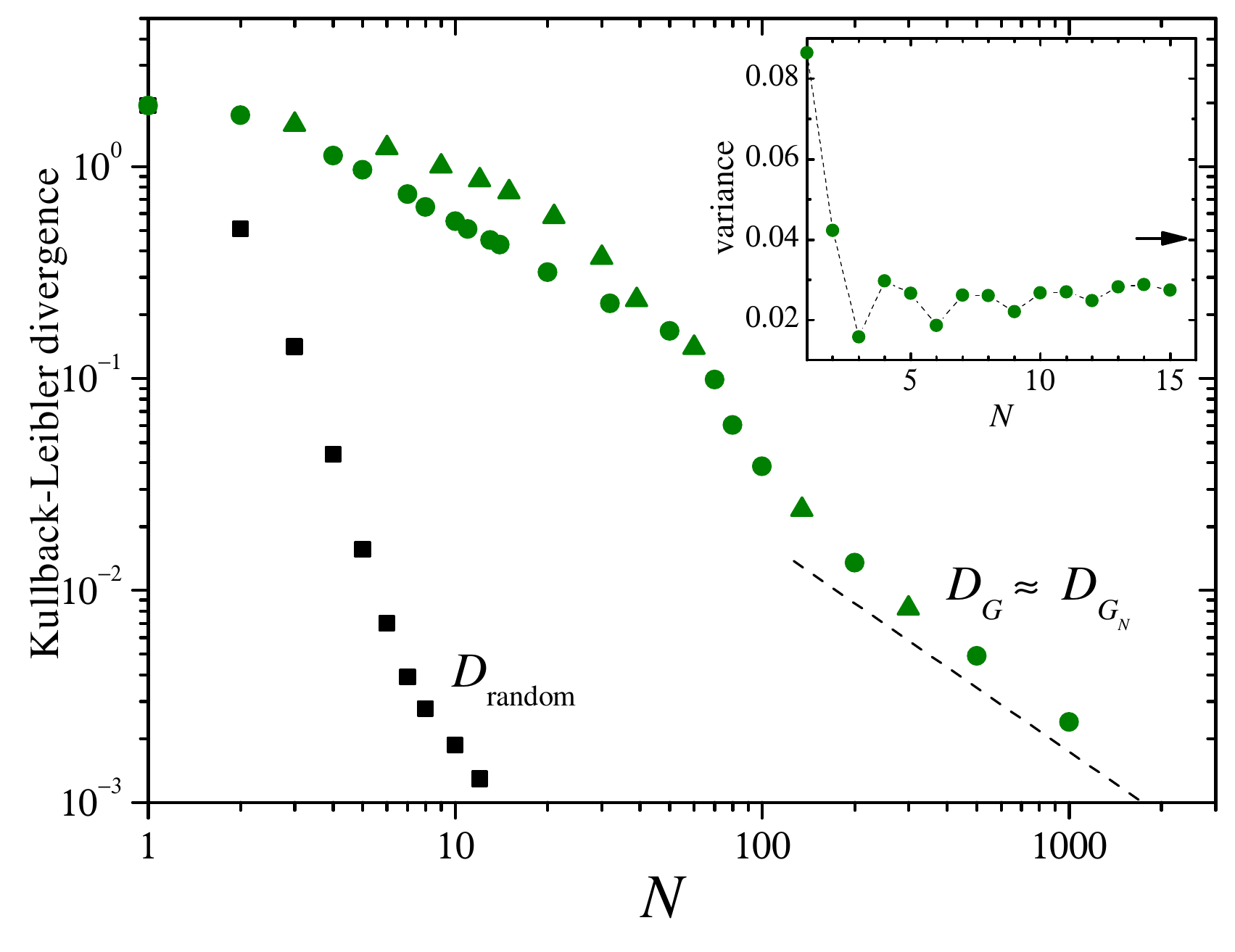}
\caption{The Kullback-Leibler divergences $D_G$ for the logistic map (\ref{logistic}) in the intermittent regime, $\lambda=3.828$, and $D_{\rm random}$, as functions of $N$. For the former, triangles correspond to values of $N$ which are multiples of $3$. The slope of the dashed straight line is $-1$. Inset: Numerical results for the variance $\sigma_{s_N}^2$ of the sums $s_N$, as a function of $N$. The arrow to the right indicates the variance obtained for large $N$. Symbols are connected by dashed lines to facilitate visualization. \label{fig07}} 
\end{figure}

Colored symbols in the main panel of Figure \ref{fig07} stand for $D_G$ in the case of the logistic map with $\lambda=3.828$, as a function of $N$. As with full chaos (cf. Figure \ref{fig05}), two distinct decay regimes are identifiable. Moreover, the behavior for $N\lesssim 50$ now contains signatures of the pseudo-periodic nature of the mapping in the ``laminar'' intervals, namely, the relatively large values of $D_G$ when $N$ is a multiple of $3$ (triangles). In fact, for those values of $N$, the distributions $\rho_{s_N}$ are narrower and sharper than for the remaining values, giving rise to higher KLDs. This is clearly illustrated by the dependence of the variance $\sigma_{s_N}^2$ on $N$, shown in the inset of the figure. After an abrupt initial decay, $\sigma_{s_N}^2$ displays oscillations of period $3$, which progressively damp out as $N$ grows. For $N\gtrsim 50$, the difference in $D_G$ for multiples of $3$ rapidly smooths out, as the KLD enters a regime where it decays approximately as $N^{-1}$, as indicated by the dashed segment of slope $-1$.

For this case of intermittent dynamics, we have also calculated $D_{G_N}$, finding qualitatively the same behavior as for $D_G$. As a matter of fact, $D_G$ and $D_{G_N}$ typically differ from each other in just about a $10$ \%. Thus, for the sake of clarity, the numerical estimations of $D_{G_N}$ were not included in Figure \ref{fig07}. As for the KLD of the distribution of random-sampling sum, $D_{\rm ramdom}$, the results of Appendix B indicate that it should decay as $N^{-1}$ for large $N$. This behavior, however, has not yet been reached in the range of values displayed in Figure \ref{fig07}. Assuming nevertheless that this is the asymptotic dependence of $D_{\rm random}$, our results suggest that the KLD  for random-sampling sums is no less than three orders of magnitude smaller than $D_G$ for large $N$.  

In summary, both for $\lambda=4$ and $3.828$, the main difference between the statistics of the sums $s_N$ obtained from the iteration of the logistic map and from a random sampling of the corresponding invariant measures, as $N$ grows, resides in their disparate rates of approach towards the asymptotic distribution. Within the range of $N$ considered in our numerical calculations, the decay of $D_G$ as $N^{-1}$ can be qualitatively understood by the lack of symmetry in the invariant measures although, strictly speaking, the corresponding result in  Appendix \ref{appB} holds for random sampling only.

Both when $\lambda=4$ and $3.828$, for $N\gtrsim 20$, the difference between $D_G$ and $D_{\rm random}$ is well above two orders of magnitude. In the intermittent case, moreover, the pseudo-periodic character of the ``laminar'' dynamics reveals itself in the form of oscillations in $D_G$ for small $N$, which are naturally absent in $D_{\rm random}$. Plausibly, pseudo-periodicity is also responsible for the slow decrease of $D_G$ during the oscillatory regime. Intermittency degrades the mixing properties of the mapping since, during the pseudo-periodic intervals, the dynamics only explore a reduced portion of the available range in $x$.

\section{A fat-tailed invariant distribution} \label{fat}

Much like the standard CLT, the CLT for deterministic systems can be generalized to the situation where the variance of the relevant variable $x$ diverges \cite{CLT3}. In particular, this is the case of invariant distributions with a sufficiently slow algebraic decay for large $|x|$: $\rho_x(x) \sim |x|^{-\alpha-1}$ with $0<\alpha<2$. Under the same hypotheses of ergodicity and aperiodicity stated in Section \ref{CLT}, and assuming for simplicity that $\bar x=0$ --for instance, due to the symmetry of $\rho_x(x)$ around zero--  the distribution of the sums 
 \begin{equation} \label{sNa}
s_N (t) =\frac{1}{N^{1/\alpha}} \sum_{k=1}^N   x(t+k-1)
\end{equation} 
[cf. Equation (\ref{sN})] converges to a stable distribution given by the Fourier antitransform of $Q_{\gamma_s} (k) = \exp \left(  -\gamma_s^\alpha |k|^\alpha\right)$, for some value of the dispersion parameter $\gamma_s$. The result for distributions with finite variance is reobtained in the limit $\alpha =2$, with $\gamma_s \equiv \sigma_s$ as defined in Equation (\ref{GK2}).

In this section, we give an example of convergence toward a stable distribution different from a Gaussian in the case of a map with a fat-tailed invariant distribution decaying as $|x|^{-2}$ for large $|x|$ (i.e. $\alpha=1$). This specific case has the analytical advantage that the stable distribution predicted by the CLT can be explicitly written out, namely, 
 \begin{equation} \label{Cauchian}
C_{\gamma_s} (s) = \frac{1}{\pi} \frac{\gamma_s}{\gamma_s^2+s^2},
\end{equation} 
which is nothing but the Cauchy (or Lorentzian) distribution. Like the Gaussian, the Cauchy distribution is a maximum-entropy distribution, but with a different constraint.

To get a deterministic chaotic map with a variable distributed following a fat-tailed function, we use the ad hoc procedure of applying a suitable transformation to a map whose invariant distribution is known in advance. Specifically, we take the Bernoulli map of Equation (\ref{Bernoulli}) with $m=2$, for which we know that the invariant distribution is the function given by Equation (\ref{rhoB}), and introduce a change of variables that transforms this function into the desired fat-tailed profile. This is formally achieved by defining the two-variable map    
 \begin{equation} \label{Cauchy}
\left\{ 
\begin{array}{l}
     u(t+1)=\{ 2u (t)\},\\
     x(t+1)=\tau [\{ 2u (t)\}],
\end{array}
\right.
\end{equation} 
where
 \begin{equation} \label{tau}
\tau (u) = \left\{ 
\begin{array}{ll}
     (2u-1)/2u & \mbox{for $0< u \le 1/2$},\\  
     (2u-1)/2(1-u) & \mbox{for $1/2\le u < 1$}
\end{array}
\right.
\end{equation} 
transforms a variable $u$ with uniform distribution in $(0,1)$ into a variable $\tau \in (-\infty,\infty)$ with distribution $\rho_\tau (\tau)= 1/2(1+|\tau|)^2$. By construction, thus, the invariant measure of variable $x$ in map (\ref{Cauchy}) is
 \begin{equation} \label{rhoC}
\rho_x (x) = \frac{1}{2 (1+|x|)^2},
\end{equation} 
with $x$ varying from $-\infty$ to $\infty$.

By analyzing the behavior of the Fourier transform of $\rho_x(x)$ near the origin, it is possible to obtain the dispersion parameter for the Cauchy distribution of sums of independently chosen values of $x$, which turns out to be $\gamma_s=  \pi/2$.  Unfortunately, the value of $\gamma_s$ when the summands are successive iterations of $x$ in map (\ref{Cauchy}) cannot be found analytically in an explicit way. However, we have numerically found that,  for $N\to \infty$, the dispersion parameter again coincides with $\gamma_s=  \pi/2$ to a high precision. This is the value of $\gamma_s$ that we use to compute the KLD $D_{C}=D(\rho_{s_N}|| C_{\gamma_s})$ between the distribution of the sums $s_N$ of Equation (\ref{sNa}) and the Cauchy distribution (\ref{Cauchian}). In addition, we do not have a practical procedure to assign a value to the dispersion parameter when the number of summands $N$ is finite. Therefore, in the present case, we do not calculate a quantity analogous to the KLD $D_{G_n}$ of Sections \ref{sec3} and \ref{sec4}.  Regarding $D_{\rm random}$, due to the non-analytic behavior of the Fourier transform of $\rho_x (x)$ at the origin, it is now not possible to use the procedure of Appendix \ref{appB} to predict how this KLD decreases as $N$ grows. Our analysis must thus rely on numerical results. 

\begin{figure*}[ht]
\includegraphics[width=10cm]{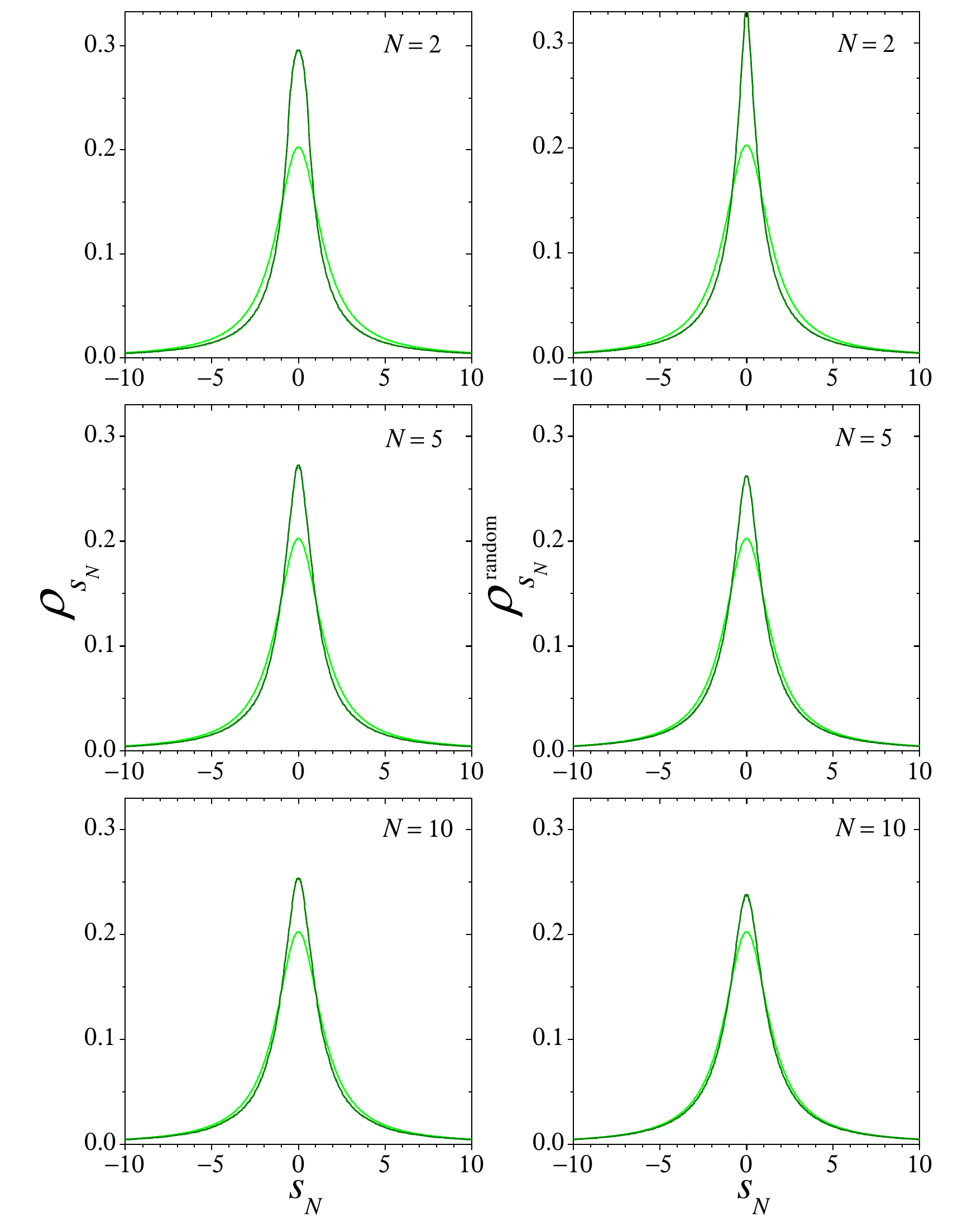}
\caption{As in Figure \ref{fig04}, for the sums of Equation (\ref{sNa}) with $x(t)$ obtained from map (\ref{Cauchy}). Note that the scales are the same in all plots.  \label{fig08}} 
\end{figure*}

\begin{figure}[ht]
\includegraphics[width=8 cm]{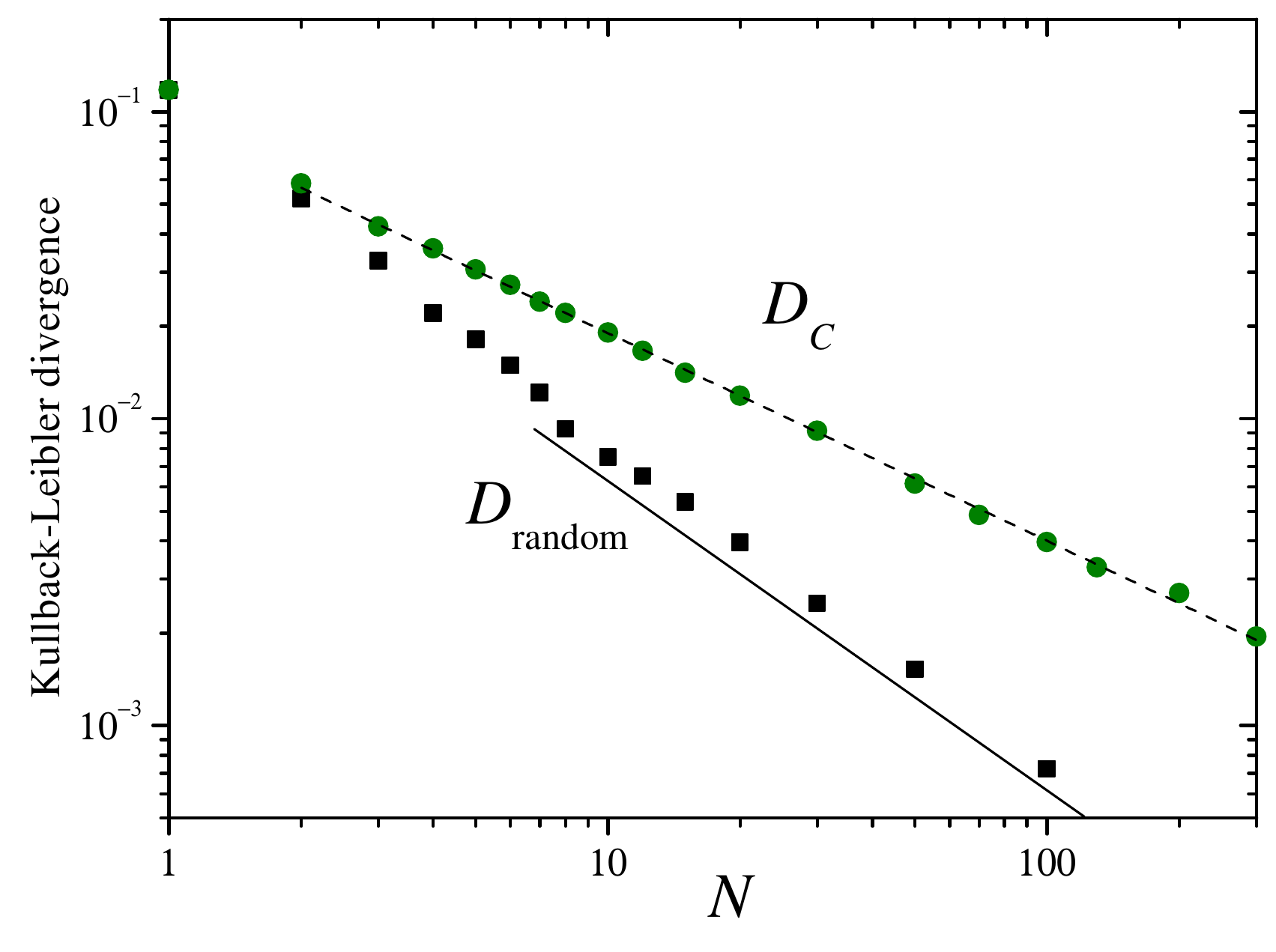}
\caption{The Kullback-Leibler divergences $D_C$ and $D_{\rm random}$ for the distributions of the sums of Equation (\ref{sNa}), with the values of $x$ obtained from the map (\ref{Cauchy}) and the distribution of Equation (\ref{rhoC}), respectively. The full straight line has slope $-1$, and the dashed line, with slope $-0.68$, is a linear fitting of $D_C$ for $N\ge 2$. \label{fig09}} 
\end{figure}

In Figure \ref{fig08}, we show the distributions $\rho_{s_N} (s_N)$ (left column) and $\rho_{s_N}^{\rm random} (s_N)$ (right column) for three small values of $N$. Light-colored curves correspond to the expected asymptotic Cauchy distribution, given by Equation (\ref{Cauchian}) with $\gamma_s=\pi/2$. Note that for $N=2$, due to the peak at $s_N=0$, the difference between $\rho_{s_N}^{\rm random}$ and the asymptotic distribution seems to be larger than that of $\rho_{s_N}$. The KLDs, however, reveal that  $\rho_{s_N}^{\rm random}$ is slightly closer to the Cauchy distribution (see Figure \ref{fig09}). For $N=10$, it is already clear that the approach to the Cauchy distribution is faster for the random-sampling sums.   Comparison with the results for the Bernoulli and the logistic maps  (cf. Figures \ref{fig01} and \ref{fig04}) suggest however that, in the present situation, the convergence to the corresponding asymptotic distribution is considerably slower than for those cases.    

Figure \ref{fig09} presents numerical results for the KLDs $D_{C}$ and $D_{\rm random}$. In order to have significant statistics in the construction of the $1000$-column histogram that represents $\rho_{s_N}(s)$ from $10^7$ samples of the sums $s_N$, we have cut off the interval of variation of $s_N$ to ($-10,10$), disregarding samples outside that interval. Otherwise, for the fat-tailed distributions involved in the present case, the calculation of the KLDs would be dominated by sampling fluctuations for large values of $|s_N|$. Along most of the range of $N$  spanned by the figure, both $D_C$ and $D_{\rm random}$ exhibit rather well-defined power-law decays.  Their different exponents, however, make that they progressively diverge from each other as $N$ grows. While $D_{\rm random}$ approximately decays as $N^{-1}$, as illustrated by the full straight line of slope $-1$, a linear fitting of $D_C$ for $N\ge 2$, shown as a dashed line, points to a slower decay with a nontrivial exponent: $N^{-0.68}$.  This result suggests that the convergence to an asymptotic distribution for the sums $s_N$ in the case of fat-tailed invariant measures may generally be characterized by unusual exponents in the decay of the KLD. This conjecture will be thoroughly explored in future work, by both analytical and numerical means.  

\section{Conclusion}

\label{conclu}

We here analyzed the convergence to the asymptotic probability density distribution $\rho_s$ of a succession of distributions $\rho_{s_N}$ for a conveniently scaled sum of $N$ samples obtained from iterations of a deterministic map. Previous analytical studies had established that a modified version of the central limit theorem (CLT) exists for these cases. Yet, as far as we know, the convergence to the limit had not been characterized. Here, we studied several archetypal examples that expose a variety of ways the limiting distributions are approached.  

Our characterization was based on the behavior of the Kullback-Leibler divergence (KLD) $D_G$ between $\rho_s$  and $\rho_{s_N}$, in that specific order. With this choice, the KLD equals the number of extra bits required to encode a sample from $\rho_s$ if the code has been optimized for $\rho_{s_N}$. The CLT for sums of random samples with finite variance predicts a KLD $D_{\rm random}$ that decreases as $N^{-2}$ if each sample is drawn from a distribution that is symmetric around its mean value, and as $N^{-1}$ if it is not. This is a bold statement, since an infinitesimal modification may suffice to turn a symmetric distribution into an asymmetric one, so even a minute modification would suffice to change the entire asymptotic behavior of the KLD --the change, however, would only become relevant at increasingly larger values of $N$, as the asymmetry tended to disappear. We are not aware of an analogous theoretical prediction for the case of correlated samples, but the results presented have revealed similar behaviors: $D_G$ decreased as $N^{-2}$ for the Bernoulli map, for which the sums are distributed symmetrically around their mean value, and as $N^{-1}$ for the logistic map, where the distributions are asymmetric. 

In both the Bernoulli and the logistic map, the rates at which $\rho_{s_N}$ approached the asymptotic distribution increased with the  strength of mixing. Moreover, for the intermittent logistic map, where mixing is virtually absent during pseudo-periodic intervals, convergence to the asymptotic distribution was particularly slow.  Therefore, even though all the explored examples were equally deterministic, their behavior differed considerably. Details in the chaotic dynamics are crucial to the behavior of $D_G$ for large $N$.

The convergence of $\rho_{s_N}$ in the Bernoulli map could be divided into two stages, one in which the distribution acquired an approximately Gaussian profile, and a subsequent one, in which the variance was adjusted to approach its asymptotic value. Remarkably, in the second stage and for sufficiently large $N$, the divergence $D_G(N)$ was equal to $D_{\mathrm{random}}(\alpha N)$ with $\alpha \approx 3.74$, implying that each sample of the deterministic map was as informative about the asymptotic distribution as $\alpha$ random samples. This equivalence could not be established in the other explored examples, since in all of them, $D_{\mathrm{random}}$ and $D_G$ decreased with $N$ with different power laws. No rescaling procedure, hence, could transform one into the other.

The last example involved variables with divergent variance. In this case, the derivation of Appendix \ref{appB} is no longer valid, and no theoretical formulation describing how $D_{\mathrm{random}}$ tends to zero is known to us. Our numerical explorations revealed a behavior proportional to $N^{-1}$ for $D_{\rm random}$, even for samples drawn from distributions that are symmetric around their mean values. The deterministic counterpart $D_G$ exhibited an even slower evolution, at a rate that is also slower than the one observed in the cases of finite variance. 

In conclusion, in all the examples explored here, the asymptotic trend of the KLD behaved as a power law. Different deterministic maps yielded different exponents, displaying a variety of behaviors. The factors that influenced the exponents were (a) the  strength of mixing in the chaotic map, (b) the tendency of the system to evolve near periodic orbits, and (c) the tails of the distribution of individual variables.  We stress that the open question of establishing a quantitative connection between the rate of mixing, on one hand, and of KLD decay, on the other, remains an interesting problem for future work. Remarkably, except for the logistic map in the intermittent regime, all the maps explored here are related to each other by simple, nonlinear transformations. Despite these deterministic functional relations, their nonlinear nature determines differences in the statistical behavior of the sums of samples drawn from each map, with a  large impact on the convergence towards their asymptotic distributions. 

\appendix \setcounter{section}{0} \setcounter{equation}{0}

\section{Calculation of the variances of $N$-term sums, Equation (\ref{GK1})} \label{appA}
\subsection*{The Bernoulli map} To calculate the variances $\sigma_{s^{(N)}}^2$ of the sums $s_t^{(N)}$ of Equation (\ref{sN}), as given by Equation (\ref{GK1}), it is necessary to compute the correlations
 \begin{equation} \label{appA1}
 \begin{split}
 c_k &=  \overline{[x(t)-\bar x][x(t+k)-\bar x] } \\
 & = \int 
  (x-\bar x)\left[ f^{(k)}(x)-\bar x\right] \rho_x(x) \mathrm{d}x \\
  & =-\bar x^2+
  \int 
  x  f^{(k)}(x) \rho_x(x) \mathrm{d}x ,    
 \end{split}
\end{equation} 
where $f^{(k)}(x)$ is the $k$-th self-iteration of $f(x)$. For the Bernoulli map, Equation (\ref{Bernoulli}), $f^{(k)}(x)$ can be explicitly given as a piecewise linear function over the interval $[0,1)$: 
 \begin{equation} \label{appA2}
   f^{(k)}(x) = m^k \left[ x-(j-1) m^{-k}\right],  
\end{equation} 
for $(j-1)m^{-k} \le x < jm^{-k}$, and $j=1,2, \dots, m^k$.  Taking into account that $\rho_x(x)=1$ over the same interval, with $\bar x=1/2$, the correlation  in Equation (\ref{appA1}) turns out to be 
 \begin{equation} \label{appA3}
  c_k =\frac{1}{12}m^{-k}.
\end{equation} 
Inserting this result in Equation (\ref{GK1}), the variances of Equation (\ref{sigmasB}) are straightforwardly obtained. Equation (\ref{appA3}) shows that, for the Bernoulli map, correlations between successive values of $x(t)$ decay exponentially with the span $k$, decreasing the faster the larger $m$ is. 

\subsection*{The logistic map in the fully chaotic regime}

The correlations $c_k=\overline{[x(t)-\bar x][x(t+k)-\bar x] }$ for the fully chaotic logistic map, $x(t+1)=4x(t) [1-x(t)]$, can be conveniently computed by exploiting the exact solution of the map \cite{Schroder,Maritz},
 \begin{equation} \label{appA4}
   x(t) =  \sin^2 \left(2^t \xi_0 \right),
\end{equation} 
where $\xi_0$ is determined by the initial condition $x(0)$ through the relation $x(0) = \sin^2 \xi_0$. Recalling that $\rho_x(x) =[\pi \sqrt{x(1-x)}]^{-1}$ and $\bar x=1/2$, Equation (\ref{appA1}) implies
  \begin{equation} \label{appA5}
 \begin{split}
   c_k  = & -\frac{1}{4}+ \\
   & + \frac{1}{\pi}
   \int_0^1 
  \sqrt{\frac{x}{1-x}}  \sin^2 \left( 2^k \arcsin \sqrt{ x} \right)  \mathrm{d}x .   
 \end{split}
\end{equation}  
The integral in this equation may look somehow intimidating but, using the change of variables $x=\sin^2 \xi$, it gets the much simpler form $2\int_0^{\pi/2} \sin^2 \xi \sin^2 (K\xi) \mathrm{d}\xi$, with $K=2^k$. Now, it can be easily shown --for instance, by induction over $K$-- that the integral equals $\pi/8$ for all integers $K>1$.  From this result, it follows that $ c_k=0$ for all $k$. Remarkably, therefore, successive iterations of the logistic map in the fully chaotic regime are  linearly uncorrelated to each other,  although their functional correlation is obviously very large. Thus, the variance of the sums $s_t^{(N)}$ is 
 \begin{equation} \label{appA6}
   \sigma_{s_N}^2 =  \sigma_x^2 = \frac{1}{8}
\end{equation} 
for all $N$.

\section{Kullback-Leibler divergence for the distribution of random-sampling sums} \label{appB} \renewcommand{\theequation}{B\arabic{equation}} \setcounter{equation}{0}

According to the Berry-Ess\'een theorem \cite{BET1,BET2}, the difference between the distribution for the sum of $N$ independent random variables and the Gaussian predicted by the standard central limit theorem decays as $1/\sqrt{N}$ or faster as $N$ grows.  We show in this Appendix that, when the distribution of the individual random variables $\rho_x(x)$ admits a cumulant expansion --i. e., when the logarithm of its Fourier transform can be expanded in powers of its variable-- that difference decays as $1/\sqrt{N}$ if  $\rho_x(x)$ is asymmetric with respect to the mean value $\bar x$, and as $1/N$ if it is symmetric. This implies that the Kullback-Leibler divergence $D_{\rm random}$ defined in the main text decays as $1/N$ in the former case, and as $1/N^2$ in the latter. In the distributions considered in the main text, the symmetry with respect to the mean value is verified for the Bernoulli map and for the logistic map in the fully chaotic regime. 

Without generality loss, we assume that the mean value over the distribution $\rho_x (x)$ of the individual random variables is zero. For the sums $s=\sum_{i=1}^N x_i/\sqrt{N}$, where $x_i$ are independent samples of $\rho_x$, the distribution $\rho_s(s)$ results from the $N$-th order self-convolution of $\rho_x(x)$. This operation is most conveniently expressed in terms of the characteristic functions (Fourier transforms) $G_x(k)$ and $G_s(k)$ of, respectively, $\rho_s(s)$ and $\rho_x(x)$. Namely, $G_s(k)= [G_x (k/\sqrt{N})]^N$, or
 \begin{equation} \label{appB1}
 \begin{split}
    \ln G_s(k) &= N \ln G_x\left( \frac{k}{\sqrt{N}} \right) \\
    & = N \sum_{j=1}^\infty \left(- \frac{ik}{\sqrt{N}} \right)^j \frac{\kappa_j}{j!},   
 \end{split}
\end{equation} 
where the sum in the right-hand side is the power expansion of $\ln G_{x} (k)$ around $k=0$, which we assume to exist, and $\kappa_j$ is the $j$-th order cumulant of $\rho_x(x)$ \cite{cum}. We recall that  $\kappa_1=\bar x=0$, $\kappa_2=\sigma_x^2$ is the variance of $x$ over $\rho_x$, and $\kappa_3 = \overline{(x-\bar x)^3}$. 

Using this information, the antitransform of $G_s(k)$ can be written as
  
\begin{eqnarray} 
  \rho_s(s) &=& \frac{1}{2\pi} \int_{-\infty}^\infty \mathrm{e}^{iks-\sigma_x^2 k^2/2}  \exp \left[ \sum_{j=3}^\infty  \frac{(-ik)^j }{N^{-1+j/2}}  \frac{\kappa_j}{j!} \right] \mathrm{d}k \nonumber \\
 &\equiv& \frac{\exp(-s^2/2 \sigma_x^2)}{\sqrt{2\pi \sigma_x^2}} + \Delta \rho_s(s), \label{appB2}
\end{eqnarray} 

\noindent
with 
 \begin{equation} \label{appB3}
 \begin{split}
   \Delta \rho_s(s) = \frac{1}{2\pi} & \int_{-\infty}^\infty  \mathrm{e}^{iks-\sigma_x^2 k^2/2}   \ \ \ \times \\ & \times \left[ \sum_{j=3}^\infty  \frac{(-ik)^j}{N^{-1+j/2}} \frac{\kappa_j}{j!} +\cdots \right] \mathrm{d}k,   
 \end{split}
\end{equation} 
where the ellipsis stands for higher-order terms in the power expansion of the second exponential in the integrand of Equation (\ref{appB2}). Note that $\Delta \rho_s(s)$ is nothing but the difference between $\rho_s(s)$ and the asymptotic Gaussian distribution $G_{\sigma_x}(s)$ and that, due to normalization, it must verify $\int \Delta \rho_s(s) \mathrm{d}s=0$. If $\rho_x(x)$ is asymmetric around zero, the third-order cumulant $\kappa_3$ is different from zero, and the leading term in powers of $N$ in $\Delta \rho_s(s)$ is given by the summand with $j=3$ in Equation (\ref{appB3}), which implies $\Delta \rho_s  \sim 1/\sqrt{N}$. If, on the other hand, $\rho_x(x)$ is symmetric around zero, we have $\kappa_3=0$. In this case,  the sum effectively starts at $j\ge 4$, and $\Delta \rho_s$ decreases as $1/N$ or faster. Of course, if $\rho_x$ is Gaussian from the start, all the higher-order cumulants vanish, and $\Delta \rho_s(s)$ is trivially equal to $0$ for all $N$.

For the Kullback-Leibler divergence we have, from Equation (\ref{KL}),
 \begin{eqnarray} 
D_{\rm random} &=& D\left(\rho_s||G_{\sigma_x} \right) \nonumber \\
&=& \int \left[ G_{\sigma_x} (s)+ \Delta \rho_s(s)  \right] \log_2 \left[ 1+ \frac{\Delta \rho_s(s)}{G_{\sigma_x} (s)}\right] \mathrm{d}s \nonumber \\
&\approx& \int \left[ G_{\sigma_x} (s)+ \Delta \rho_s(s)  \right]   \frac{\Delta \rho_s(s)}{G_{\sigma_x} (s)}  \mathrm{d}s \nonumber \\
&=& \int \frac{\left[ \Delta \rho_s(s) \right]^2 }{G_{\sigma_x} (s)} \mathrm{d}s, \label{appB4}
\end{eqnarray} 
where the approximation holds if $\Delta \rho_s(s)$ is sufficiently small. If the distribution $\rho_x(x)$ is asymmetric around zero,  since $\Delta \rho_s(s)$ decays asymptotically as  $1/\sqrt{N}$, the decay of $D_{\rm random}$ turns out to be as $1/N$. If it is symmetric, $D_{\rm random}$ decays as $1/N^2$ or faster, depending on whether the subsequent cumulants vanish or not.

%\acknowledgments{In this section you can acknowledge any support given which is not covered by the author contribution or funding sections. This may include administrative and technical support, or donations in kind (e.g., materials used for experiments).}

%\end{adjustwidth}

\end{document}